\documentstyle[12pt,psfig]{article}
\def\bi{\bibitem}
\newcommand{\be}{\begin{equation}}
\newcommand{\ee}{\end{equation}}
\newcommand{\beq}{\begin{eqnarray}}
\newcommand{\eeq}{\end{eqnarray}}
\newcommand{\bear}{\begin{array}}
\newcommand{\ear}{\end{array}}

\begin{document}
\title{The CWKB Method of Particle Production Near Chronology Horizon}
\author{S. Biswas$^{*a)}$, P. Misra $^{a)}$and I. Chowdhury$^{a)}$\\
a) Department of Physics, University of Kalyani, West Bengal,\\
India, Pin.- 741235\\
$*$email:sbiswas@klyuniv.ernet.in}
\date{}
\maketitle
\begin{abstract}
In this paper we investigate the phenomenon of particle production of massles scalar field, in a model of spacetime where the chronology horizon could be formrd, using the method of complex time WKB approximation (CWKB). For the purpose, we take two examp
les in a model of spacetime, one already discussed by Sushkov, to show that the mode of particle production near chronology horizon possesses the similar characteristic features as are found while discussing particle production in time dependent curved ba
ckground. We get identical results as that obtained by Sushkov in this 
direction. We find, in both the examples studied, that the total number of particles remain finite at the moment of the formation of the chronology horizon.  
\end{abstract}
PACS numbers(s): 04.20.Gz, 04.62.+v
\section{\bf{Introduction}}
The study of closed timelike curves (CTC) has gained a serious attention since its introduction by Morris, Throne, and Yurtsever \cite{mty:prl}. Hawking \cite{swh:pmgm,swh:prd} was of the view that such closed timelike curves are not allowed in real world
 whioch we describe through the standard laws of physics. This view now runs 
with the name, `Hawking chronology protection conjecture'. The principle of general relativity allows in its framework the occurence of closed timelike curve and has been demonstr
ated by various authors \cite{wjvs:prs,kg:rmp,ntu:jmp,cwm:rta,jrg:prl,kst:grg,mv:lw}. The arguments behind not allowing closed timelike curves are that the renormalized energy momentum tensor diverges at the Cauchy horizon (gennerated by closed null geode
sic) separating the regions with CTC from the region without closed causal curves. Now we have many examples where we find bounded renormalized stress-energy tensor near the chronology horizon \cite{svk:prd,svs1:cqg,svs2:cqg,ll:cqg,mv:plb}. In Gott \cite{
jrg:prd} we find an elaborate discussion how CTCs play role in the early universe and could be the mother of itself i.e., the universe creates itself.
\par
If we believe in semiclassical quantum gravity, the laws of standard physics allow the wave function of the universe to be in the description for which we have no satisfactory initial conditions. In this regime, the allowance of CTC might be a step furthe
r to understand the `nothing' from where the universe emerges [16]. Now we know that the particle production is the dominant factor for the creation of the matter in the universe. So, if CTC has any role in the formation of the universe it is necessary to
 investigate whether the particle production near such CTC destroys it or not. Sushkov \cite{svs:prd} tried to get an answer, considering particle production near chronology horizon and found that the total number of particles remain finite at the moment 
of the formation of the chronology horizon. We adopt here the similar approach but with the method of CWKB. In a general class of spacetime, finding of mode solutions and then the calculation of Bogolubov coefficients , to study particle production, is a 
very difficult task. The CWKB offers an way out in this direction \cite{bis:grg,bis1:cqg,bis2:cqg,bis3:grg}. In this work we apply the method of CWKB to study particle production in a spacetime with a property of having CTC at distant future. For the purp
ose we cosider a two dimensional model of spacetime akin to Sushkov [17]. The present work on the one hand substantiates the calculation of Sushkov and on the other hand allows one to use CWKB in a more general class of spacetimes with possibility of form
ing closed timelike curves. In this work we take two examples to elucidate our stand.
\par
We use the units $c=\hbar=G=1$ throughout the paper.
\section{\bf{Model of Spacetime}}
We consider a model of spacetime in which we find the chronology horizon being formed at distant future with no such behaviour at early times. We consider the metric
\be
ds^2=d\,\eta\,^2+2a(\eta)\,d\,\eta\,d\,\xi-[1-a^2(\eta)]d\,\xi^2,
\ee
where $a(\eta)$ is a monotonically increasing function of $\eta$ with the following behaviour:
\beq
a(\eta)\rightarrow 0\,\,\,\, if\,\,\,\, \eta\rightarrow -\infty,\nonumber\\ 
a(\eta)\rightarrow a_0\,\,\,\, if\,\,\,\, \eta\rightarrow +\infty,
\eeq
where $a_0$ is some constant. To effectuate the occurence of CTC we consider the strip $\{\eta \in(-\infty,+\infty),\,\,\xi\in[0,L]$ on the $\eta - \xi$ plane and assume that the points $\xi=0$ and $\xi=L$ are identified i.e., $(\eta,0)\equiv (\eta,L)$. T
his produces a manifold with the topology of a cylinder: $R^1\times S^1$.
The regularity of spacetime with $R^1\times S^1$ topology is ensured as follows. The metric coefficients do not depend upon $\xi$, so the metric and its derivative takes the same values at the points $(\eta,0)$ and $(\eta,L)$. Hence the internal metrics a
nd curvatures are identical at both lines $\gamma^-:\xi=0$ and $\gamma^+:\xi=L$.
\par 
To study particle production it is necessary to identify the ``in" and ``out" regions where the particle and the antiparticle states are defined. From (1) and (2) it is evident that at $\eta\rightarrow -\infty$, we have exactly the Minkowski form
\be
ds^2=d\,\eta\,^2-d\,\xi\,^2.
\ee
In the future, $\eta\rightarrow +\infty$ we have 
\be
ds^2=d\,t\,^2-d\,x\,^2,
\ee
where the new coordinates are
\be
t=\eta+a_0\xi,\,\,\,\,x=\xi
\ee
Thus in out-region and in-region we can construct Minkowski-like vacuum. For any point $(\eta,\xi)$ in $(R^1\times S^1)$ there are infinite number of images of points $(\eta,\xi+L)$ in the covering space, the whole $(\eta,\xi)$ plane. Thus we have the equ
ivalence relation between $R^1\times S^1$ and the covering space as
\be
(\eta,\xi+L)\equiv (\eta,\xi)
\ee
Now we obtain the equation of null curves setting $ds^2=0$ in Eq.(1). We get
\beq
\xi+\int^\eta\frac{d\,\eta\,^\prime}{1+a(\eta\,^\prime)}=const.(\equiv C^-),\\
\xi-\int^\eta\frac{d\,\eta\,^\prime}{1-a(\eta\,^\prime)}=const.(\equiv C^+).
\eeq
Here $C^-$ describes the {\it{left}}-hand branch of the future light cone and $C^+$ describes the {\it{right}}-hand branch. From the asymptotic properties Eq.(2), it follows from Eqs.(7) and (8) that for the null geodesics in the in-region we have
\be
\eta+\xi=const,\,\,\,\,\,\eta-\xi=const,
\ee
and in the out-region they are
\be
\eta+(1+a_0)\xi=const,\,\,\,\,\,\,\eta-(1-a_0)\xi=const.
\ee
Eqs. (10) now contain the clue for forming the chronology horizon. As $\eta$ gets larger the {\it{right}}-hand branch of the future light cone rotates  rightward and ultimately if $a_0\rightarrow 0$, it coincides with the $\xi$-axis so that we get $\eta=c
onstant.$ Now from Eq. (6) we find that the curve in the $\xi$ direction is closed. If this occurs at the moment $\eta=\eta^*$  such that $C^+$ branch becomes horizontal, then the closed null curves appear in our model. We now say that a time machine is b
eing formed at this moment of time. If $\eta^*=\infty\,\,\,\,(a_0=1) $, the time machine is formed in the infinitely far future. Otherwise, if we have $\eta>\eta^*$, the closed line $\eta=constant$ lies inside of the light cone. For more details, the read
er is referred to ref.[17]. The discussion exemplifies the occurece of CTC in the spacetime defined by the metric (1).
\par 
We now investigate the particle creation in such a spacetime.
\section{\bf{Particle Creation}}       
We consider the massless scalar field equation 
\be
\Box\,\phi=0
\ee
in the metric given by Eq.(1). In this metric the wave equation reads
\be
\left[(1-a^2)\partial_\eta^2+2a\partial_\eta\partial_\xi-\partial_\xi^2-2a^\prime\partial_\eta+a^\prime\partial_\xi\right]\phi(\eta,\xi)=0,
\ee
where $\partial_\eta=\partial/ \partial \eta,\,\,\partial_\xi=\partial/ \partial\xi$, and a prime denotes the dervative on $\eta$, $a^\prime=da/d\eta$. In the in-region $a(\eta)\rightarrow 0$ and $a^\prime(\eta)\rightarrow 0$ Eq. (12) reduces to the form
\be
\left[\partial_\eta^2-\partial_\xi^2\right]\phi(\eta,\xi)=0.
\ee
The complete set of solutions of this equation is
\be
\phi_n^{\pm,in}=D_n^{(in)}e^{ik_n\xi}e^{\mp i\omega\eta},
\ee
where $\omega =\vert k_n\vert$ with $k_n=\frac{2\pi n}{L},\,\,\,\,n=\pm 1,\pm 2,...$ being determined from the boundary condition 
\be
\phi(\eta.\xi+L)=\phi(\eta,\xi).
\ee
From the normalization condition
\be
(\phi_n,\phi_{n^\prime})=-i\int_0^L\,d\xi\left(\phi_n\frac{\partial\phi^*_{n^\prime}}{\partial \eta}-\frac{\partial\phi_n}{\partial\eta}\phi^*_{n^\prime}\right)_{\eta =const.}=\delta_{nn^\prime} 
\ee
we get
\be
D_n^{(in)}=\frac{1}{\sqrt{4\pi\vert n\vert}}. 
\ee
To obtain the mode solution in the out region, we have $a(\eta)\rightarrow a_0)$ and $a^\prime(\eta)\rightarrow 0$ so that Eq.(12) reduces to
\be
\left[(1-a_0^2)\partial_\eta^2+2a_0\partial_\eta\partial_\xi-\partial_\xi^2\right]\phi(\eta,\xi)=0.
\ee
We find the solution as before as 
\be
\phi_n^{(\pm,out)}=D_n^{(out)}e^{ik_n\xi}e^{ik_na_0\beta^{-1}\eta}e^{\mp \omega\beta^{-1}\eta},
\ee
with $\beta=1-a_0^2$ and
\be
D_n^{(out)}=\sqrt{\frac{\beta}{4\pi\vert n\vert}}.
\ee
It is evident from Eqs.(14) and (19) that $\omega_{in}=\omega\neq\omega_{out}=(\omega\pm k_na_0)\beta^{-1}$ so that there is possibility of particle production as the particles evolve from the in-vacuum to out-vacuum. Let us introduce
\be
\phi_n(\eta,\xi)=v(\eta)exp\left(-\int^\eta\,\,\,\frac{a(ik_n-a^\prime)}{1-a^2}d\eta\right)exp(ik_n\xi)
\ee
in Eq.(12) to obtain
\be
v^{\prime\prime}+\Omega^2(\eta)v=0,
\ee
where
\be
\Omega^2(\eta)=\frac{k_n^2+a^{\prime 2}+a^{\prime\prime}a(1-a^2)}{(1-a^2)^2}
\ee
We solve Eq.(22) only for modes for which 
\be
k_n^2>>a^{\prime 2},\,\,\,\,k_n^2>>a^{\prime\prime}a(1-a^2)
\ee
i.e., we consider only those modes whose wavelength is much less than a typical scale of variation of the function $a(\eta)$. We now consider two cases:
\beq
(I)\,\,\,\,\,\,a(\eta)=\frac{a_0^2}{1+exp(-2\gamma \eta)},\\
(II)\,\,\,\,\,\,a(\eta)=\frac{1}{2}a_0^2(1+tanh\gamma \eta)
\eeq
The case (II) has already been discussed by Sushkov [17]. We will now use the method of CWKB to obtain the number of created particles near the chronology horizon where $a_0\rightarrow 1$. According to CWKB, the boundary conditions for particle production
 is taken as
\beq
v(\eta\rightarrow -\infty) & \sim & e^{iS(\eta,\eta_0)},\\
v(\eta\rightarrow +\infty) & \sim &  e^{iS(\eta.\eta_0)}+ R\,\,e^{-iS(\eta.\eta_0)}.
\eeq
Here we have neglected the WKB pre-exponential factor for convenience and
\be
S(\eta,\eta_0)=\int_{\eta_0}^\eta\,\,\Omega(\eta)d\,\eta
\ee
and the reflection amplitude $R$ that accounts particle production is given by
\be
R=\frac{-ie^{2iS(\eta_1,\eta_0)}}{\sqrt{1+e^{2iS(\eta_1,\eta_2)}}}
\ee
where the turning points are given by the condition $\Omega(\eta_{1,2})=0$. In (30) the denominator takes into account the the repeated reflections between the turning points $\eta_1$ and $\eta_2$. In case of single turning point, we have
\be
R=-ie^{2iS(\eta_1,\eta_0)}
\ee
In above equations $\eta_0$ is an arbitrary real point and does not affect the magnitude of $R$.  The derivation of Eq.(30) can be found in our earlier works [18,19,20]. For case (I) we are to evaluate the integral
\be
I=S(\eta,\eta_0)= k_n\int^\eta \frac{d\eta}{1-\frac{a_0^2}{1+exp(-2\gamma\eta)}}.
\ee
We avoid the $\eta_0$ limit as it would give a real contribution to $I$ and hence does not contribute to $|R|$ in Eq.(30). We now substitute $\tau=exp(-2\gamma\eta)$ and take $1-a_0^2=\beta$ in the above integral and find
\be
I=-\frac{k_n}{2\gamma}\int^\eta\frac{(\tau+1)d\,\tau}{(\beta\tau+1)\tau}.
\ee
We now have the turning point at $\tau=-1$ i.e., at complex $\eta=\pm i\pi/{2\gamma}$. After evaluation we get
\be
I=-\frac{k_n}{2\gamma}[ln\,\tau+\frac{1-\beta}{\beta}(ln\tau-ln(\beta+\tau))]
\ee
To evaluate $I$ we should be careful. For $\beta\neq 0$, only the first and second term contributes but for $\beta\rightarrow 0$, all the three terms contribute and we get
\be
 I=-\frac{k_n}{2\gamma}i\pi.
\ee
Using the expression of $R$, we now get
\be
\vert R \vert^2=e^{-\frac{\pi k_n}{\gamma}}
\ee
The Bogolubov coefficient $\beta_n$ is now evaluated as [18]
\beq
\vert\beta_n\vert^2 & = &\frac{\vert R\vert^2}{\vert T\vert^2}\nonumber\\
& = & \frac{\vert R\vert^2}{(1-\vert R\vert^2)},
\eeq
so that
\be
\vert\beta_n\vert^2=\frac{e^{-\pi k_n/{\gamma}}}{1-e^{-\pi k_n/ \gamma}}
\ee
\par Let us now cosider the case II. Here we are to evaluate the integral
\be
I=k_n\int\frac{d\,\eta}{1-\frac{1}{2}a_0^2(1+tanh\gamma\eta)}
\ee
As before we put $t=e^{2\gamma\eta}$ and get
\be
I=\frac{k_n}{2\gamma}\int\frac{dt(t+1)}{(\beta\,t+1)t},
\ee
where $\beta=1-a_0^2$. After evaluating the integral we find
\be
I=\frac{k_n}{2\gamma}ln\,t-\frac{k_n(\beta-1)}{2\gamma\beta}ln(\beta\,t+1).
\ee
 Here the turning point is again at $t=-1$ so that
\be
I(t=-1)=\frac{i\pi k_n}{2\gamma}+\delta
\ee
Here $\delta$ term comes from the real point $\eta_0$ as well as from the real part contribution of the above integral and it does not contribute in the expression of $\vert R\vert$. Hence in both the cases we find the number of particles produced in the 
mode $k_n$ is
\be
N_n=\vert\beta_n\vert^2=\frac{e^{-\pi k_n/\gamma}}{1-e^{-\pi k_n/\gamma}}.
\ee 
We may now conclude that the total number of particles $N=\sum_n\,N_n$ will be finite because the spectrum (43) is exponentially decreasing. In obtaining (43) we have used the condition (24) which when evaluated reduces to
\be
k_n^2>>\,n\frac{a_0^2}{T^2}\,,
\ee
where $T=(2\gamma)^{-1}$ gives the typical time variation of the function $a(\eta)$ from one asymptotical value to another one and $n$ is a factor of $O(1)$.   \section{\bf{conclusion}}
The chronology horizon is formed when $\beta=1-a_0^2\rightarrow 0$ i.e., at the moment of time $\eta^*\,<\infty$ if $a_0>1$. In the case  $a_0=1$, closed lightlike curves are formed in the infinitely far future. In our work we obtain almost identical resu
lts as that obtained by Sushkov and substantiate the conclusion that the phenomenon of particle creation could not prevent the formation of a time machine. For $a_0<1$ we have no causal pathologies.
\par
While discussing the particle production in curved spacetime we noticed that in CWKB we do not require to know the exact solutions of the problem in question. The vacuua are the WKB vacuua as defined by Parker. As the particles evolve from the in-vacuum t
o the out-vacuum they find $\vert 0>_{in}\neq \vert0>_{out}$. The reasons for such a change is that the particle moves into the Euclidean vacuum; in otherwords, it encounters complex $\eta$ or $t$ plane and causes the instability of vacuum causing particl
e production. From the discussion it is clear that since $\Omega\neq 0$ in the real time plane we are basically studying the over the barrier reflection. As $(\frac{da}{d\eta})_0\sim \gamma$ is in the expression in the denominator of $Im\,\,I$, a sharp ri
se is thus needed so that $exp(-\pi k_n/\gamma)$ remains small. What we observe that the particles so produced still survive when the chronology horizon is formed in distant future. It is therefore essential to investigate how would this energy density of
 produced particles affect the spacetime metric and the formation of chronology horizon.
\begin{center}
{\bf{Acknowledgements}}
\end{center}
The authors would like to thank Dr. Bijan Modak for helpful discussions during the preparation of the manuscript.\\
    
\end{document}